\documentclass[twocolumn,showpacs,preprintnumbers,amsmath,amssymb]{revtex4}
\usepackage{graphics}
\usepackage{dcolumn}
\usepackage{bm}

\begin{document}

\preprint{APS/123-QED}

\title{Exploring Parameter Constraints on Quintessential Dark Energy: the Albrecht-Skordis model}

\author{Michael Barnard}
\author{Augusta Abrahamse}
\author{Andreas Albrecht}
\author{Brandon Bozek}
\author{Mark Yashar}
 \affiliation{University of California, Davis.}

\begin{abstract}
We consider the effect of future dark energy experiments 
on ``Albrecht-Skordis'' (AS) models
of scalar field dark energy using the Monte-Carlo Markov chain method.
We deal with the issues of parameterization of these models, and have
included spatial curvature as a parameter, finding it to be important.
We use the Dark Energy Task Force (DETF) simulated data to represent
future experiments and report our results in the form of likelihood contours in the chosen
parameter space.  Simulated data is produced for cases where the
background cosmology has a cosmological constant, as well as cases
where the dark energy is provided by the AS model. 
The latter helps us demonstrate the power of DETF Stage 4 data in the
context of this specific model. Though the AS model can produce
equations of state functions very different from what is possible with
the $w_0-w_a$ parametrization used by the DETF, our results are
consistent with those reported by the DETF.  
\end{abstract}

\maketitle

\section{Introduction}
Current astronomical observations are best fit by cosmologies
containing a significant density of energy with negative pressure.
This has been dubbed dark energy, and its exact nature remains a
mystery.  Though the properties of the dark energy are poorly
constrained by current data, a number of proposed experiments will
have the reach to probe them.  Most assessments of proposed
experiments use abstract parameters such as the ``$w_0-w_a$''
parametrization of the dark energy equation of state used by the Dark Energy Task
Force (DETF) \cite{Albrecht:2006um},
or generalized forms of scalar field potentials
\cite{huterer:081301,huterer:123527,huterer:083503}.
In order to fully evaluate future
experiments it is useful to understand the impact and discriminating
power they will have on specific proposed models of dark energy.
In this paper, we will focus
on one scalar field (or ``quintessence'') model of dark energy,
the so-called ``Albrecht-Skordis'' (AS) model \cite{Albrecht:1999rm,Skordis2000}.

 The AS model is interesting for a number of reasons. Fields of the
 form needed for the AS model can arise as radius moduli of curled
 extra dimensions \cite{Albrecht:2001xt}, and, while these moduli may
 be treated as scalar  fields in gravitational calculations, they are
 not subject to the quantum corrections that can create serious
 problems for most  quintessence models \cite{Carroll98,Kaloper:2005aj}.
The phenomenology
 of these models is also interesting because, unlike the majority of
 dark energy models, the AS quintessence field can contribute a
 significant fraction of the total cosmic energy density {\em
 throughout} the history of the universe.

Central to this work are the projected data sets (or ``data models'')
created by the DETF\cite{Albrecht:2006um}.
We use the 
supernova, weak lensing, baryon oscillation, and Planck data sets
(though not the cluster data sets, for technical reasons similar to
those outlined in \cite{Albrecht:103003}), to show how
these data sets would impact the range of possible parameters for
quintessence models.
To this end, we use Monte-Carlo-Markov chains (MCMC) on these data sets
to analyze dark energy and cosmological parameters,
rather than the fisher matrix methods used by the DETF.
In addition to using ``standard'' simulated data based on a universe
with a cosmological constant, we create projected data sets based
on an AS model universe, and run the MCMC around these to highlight the
discriminating power of the projected data sets.
This paper is one in a series of papers that consider
different scalar field models using similar
methods\cite{AbrahamseMock}. (Technical information about our data models and
MCMC methods will be presented in an appendix of another paper in the
series \cite{AbrahamseMock}).

We show that the stage-by-stage improvement in parameter
constraints for the Albrecht-Skordis model is similar to the relative
improvement in
the $w_0-w_a$ constraints seen by the DETF.  We do the same
for data sets generated around a specific AS model which shows
small deviations from $w(a)=-1$ in the present epoch. This AS model is
chosen to illustrate a case where if the real universe is described by
an AS model, pure cosmological constant dark energy can be ruled out by
a large margin using good Stage 4 data. We also note some features in
the AS parameter contours when fitting to cosmological constant based
data that reflect the fact that the AS models we consider can only ever
duplicate a cosmological constant in an approximate manner.

\section{Parameter space of the Albrecht-Skordis model}

\begin{figure}
\centering
\scalebox{0.45}{\includegraphics{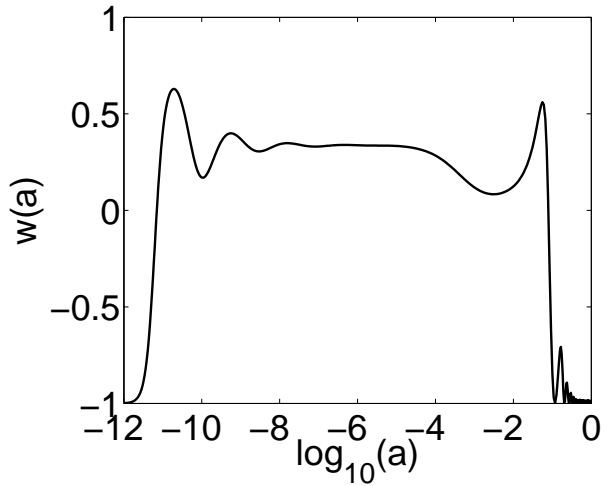}}
\caption{The equation of state $w$ for the scalar field of a typical
  AS model.  Note that here the $a$ scale is logarithmic in order to
  show behavior on all time scales. When the background energy density
  drops to a level approaching the initial energy of the field during
  radiation domination, the energy density goes through a transient
  before mimicking the equation of state of radiation ($w=1/3$).  At matter
  domination, the field again goes through a transient, but the field
  approaches the local minimum in the potential and becomes the dominant form of 
  energy (with $w \rightarrow -1 $) before it can stabilize with a
  mater-like equation of state ($w=0$). } 
\label{fig:wofloga30}
\end{figure}

Scalar field models of dark energy, or quintessence, are considered in
the framework of an FRW cosmology, with and equation of motion of
\begin{equation}
\ddot{\phi}+2\frac{\dot{a}}{a}\dot{\phi}+a^2\frac{\partial V}{\partial \phi}=0
\end{equation}
where $a$ is the scale factor, $\phi$ is the homogeneous scalar field,
and $V$ its potential.
The background is assumed to be the homogeneous Friedmann equation
\begin{equation}
  H^2=\frac{8\pi G}{3}\left(\rho_r+\rho_m+\rho_\phi\right)-\frac{k}{a^2}
\end{equation}
The equation of state parameter of the scalar field is
\begin{equation}
w=\frac{p_\phi}{\rho_\phi}
\end{equation}
where $p_\phi$ and $\rho_\phi$ are the pressure and  density of the
scalar field.  Non-relativistic matter has $w=0$, radiation has
$w=1/3$, and a cosmological constant has $w=-1$; a homogeneous scalar field
can have any behavior in this range, depending on the actual time
evolution $\phi(t)$.
The Albrecht-Skordis model\cite{Albrecht:1999rm} postulates a scalar field with an
``exponential-with-pre-factor'' potential,
\begin{equation}
V(\phi)=V_0\left[\left(\phi-B\right)^2+A\right]e^{-\lambda\phi}
\label{eq:ASo}
\end{equation}
One of the attractions of this model is that realistic cosmologies can
be achieved for cases where the parameters in this potential are all
roughly of order unity in Planck units. Specifically, with such
parameter choices the potential can have a local minimum with a
height consistent with the dark energy density observed today.  It is
assumed that the field starts high up
on the potential. The field will then rapidly approach an attractor
solution where 
the field  mimics the equation of state of the dominant energy.  It
continues this tracking
behavior until the field reaches the (approximately) quadratic shaped local minimum,
at which point it begins a damped oscillation around that minimum,
giving its equation of state a wave like variation as the field becomes the
dominant energy.  Figure \ref{fig:wofloga30} shows $w(a)$
throughout the history of the universe for a typical AS model.

\begin{figure}
\centering
\scalebox{0.45}{\includegraphics{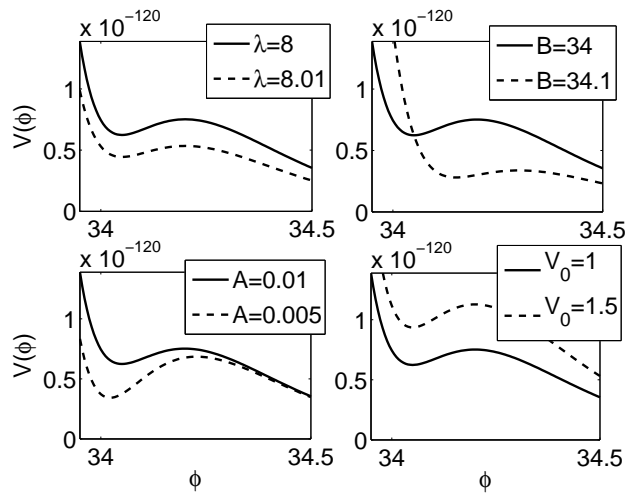}}
\caption{The effects of the original parameters (from Eqn. \ref{eq:ASo}) on the
potential minimum. All measures are in Planck units.  Note that very small changes in $\lambda$ or $B$
cause substantial changes in the height of the minimum.}
\label{fig:ASoldpars}
\end{figure}

The potential of the AS model has in principle four
degrees of freedom important to cosmology, most of which are
complicated combinations of the various parameters.  Figure
\ref{fig:ASoldpars} illustrates some of the parameter dependences of
the potential.  Perhaps the most important degree of 
freedom, because it determines the scalar field fraction of the
total energy density during the tracking behavior, is the logarithmic
slope, $(lnV)'$ or $\frac{V'}{V}$; this is almost entirely controlled by $\lambda$.
The height of the local minimum sets the
dark energy density today.  The curvature of the potiential at its minimum
is related to the height of the local maximum.
The height of the local maximum determines whether the field stops in the
minimum or rolls off to infinity, and the curvature at the minimum sets the
frequency of the late time oscillations.  We have found that the
oscillation frequency is not measurable by any of the simulated data
sets we considered.   Also, the parameter space for models that roll
off to infinity but still give realistic cosmic acceleration is a only
an exponentially thin and extremely finely tuned region\cite{Barrow:2000nc}. We choose to ignore
this exotic behavior and focus completely on parameters for which the
field (classically) stays in the local minimum.

\begin{figure}
\centering
\scalebox{0.45}{\includegraphics{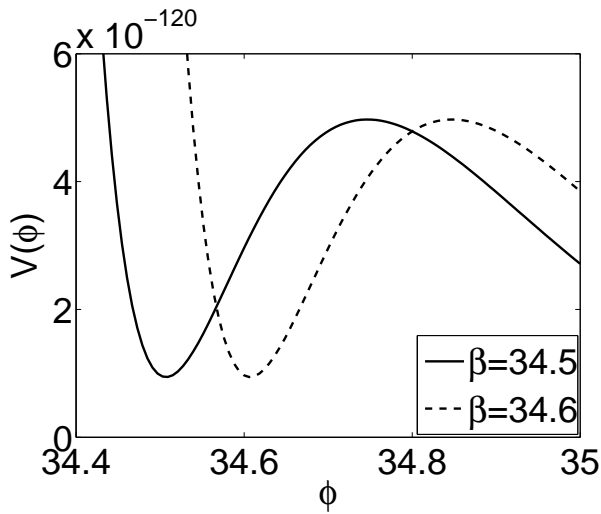}}
\caption{Under the new parameterization, $V_0$ maintains it's previous effects, while the new parameter $\beta$ becomes the primary parameter for the AS model.  $\beta$ sets the exponential slope as well as the minimum curvature.}
\label{fig:ASnewpars}
\end{figure}

It will be helpful to re-express Eqn. \ref{eq:ASo} in the following parameters:
\begin{equation}
V(\phi)=V_0\left[\chi\left(\phi-\beta\right)^2+\delta\right]e^{-\lambda\phi}
\end{equation}
Because the number of parameters is greater than the number of
parameters measured by our simulated data the AS model has a number
of degenerate directions. 
Along these degenerate directions, the
parameters can change without significantly affecting the observables.
The most apparent of these directions are the ways that the height of
the local minimum of the potential can be changed.  This height is
exponentially sensitive to the product of the exponential factor $\lambda$,
 and $\beta$, which primarily controls the location of the
minimum. This degeneracy can be stabilized by fixing the product of
$\lambda$ and $\beta$, so that $\lambda=\frac{272}{\beta}$. There is
also a dependence of the minimum on $\delta$ and
$\frac{\lambda^2}{\chi}$.  To work around these difficulties, we
fix $\delta$, and vary $\chi$ as a function of $\beta$ so that the overall constant $V_0$ 
is the only parameter that adjusts the height of the
minimum.  This necessarily excludes the portion of the
parameter space where there is no minimum (with essentially no loss of
generality, as discussed above). Also, these parameter constraints do
not allow the oscillation frequency near the local minimum to vary
independently of the other parameters. This feature is required to
successfully run the MCMC calculations since none of the data sets are
sensitive to the oscillation frequency.  Figure \ref{fig:ASnewpars} shows how
$\beta$ affects the new parameterization.

Aside from degeneracy in the specifics of the potential, there 	are
broad patches of parameter space that are, for most data sets,  not
detectably different from a cosmological constant. This happens when
the exponential slope is steep, and the field slides very early down
to the local minimum and oscillates in a shallow and rapid manner, resulting
in an equation of state very nearly $-1$ at the times of interest.

The starting value of the scalar field is mostly unimportant in the AS
model, but it should be noted that there are values of $\beta$ and the
initial $\phi$ that can cause problems.  For large values of $\beta$,
if the scalar field starts at a point higher on its potential than the
radiation density during nucleosynthesis, then the scalar field will
form a significant fraction of the energy density at that epoch.  Such regions of
parameter space would be ruled out by observations of primordial
element abundances, though our algorithm does not contain such
considerations.

Other related problems with the algorithm specific to
the AS model include the last scattering surface.  The algorithm does
not calculate the size of perturbations at the time of last
scattering, but takes them as an input, so the effect of the scalar
field on that size is not taken into account by the algorithm even
when the scalar field energy density is significant.  With respect to
the growth of linear density perturbations, the algorithm does not use
the scalar field in its calculation until a redshift of ten.  The
effect of these issues is that some constraining power has been
ignored.
Albrecht and Skordis investigated this issue \cite{Skordis2000} to
some degree, but while we do not believe this affects the conclusions
of this paper, this matter remains interesting and worthy of further exploration.

In our parameterization, the amplitude of the oscillations in the equation of state at
late times is primarily determined by the value of $\beta$.
For small values of $\beta$ the oscillation amplitudes are very small,
but become large for values of $\beta$ approaching 100.  This
somewhat unintuitive behavior can be explained by noting
that, when the dominant form of energy is scaling as $a^{-n}$,
$\rho_{\phi}/\rho_c =\frac{n}{\lambda^2}$ (where $\rho_c$ is the
critical density)\cite{Ferreira:1997hj}.  Thus, the smaller
$\beta$ is (and the larger $\lambda$ is), the further the
field energy is below the matter density during mater
domination, and the more Hubble friction will slow the field.
The closer the scalar field energy density is to the matter
density during matter domination, the more kinetic energy
remains in the field when it becomes dominant, and thus free
to oscillate.

\begin{figure}
\scalebox{.45}{\includegraphics{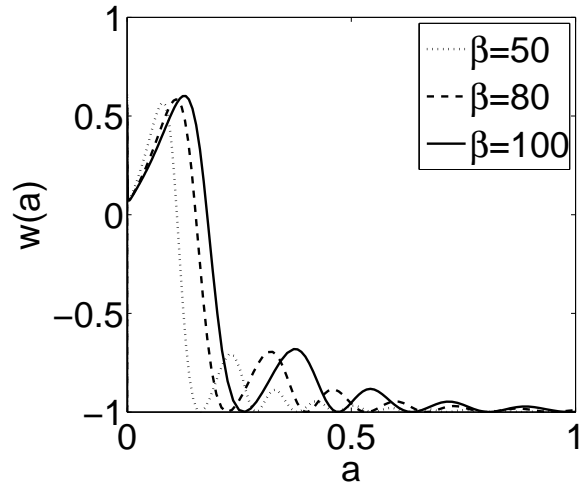}}
\caption{This figure illustrates how the primary parameter $\beta$
  affects the $w(a)$ behavior of the AS model. The AS model-based
  data was generated using $\beta=80$.}
\label{fig:wofacomb}
\end{figure}

\section{Constraining the AS model around Cosmological Constant model data sets}
\begin{figure}
\centerline{
\scalebox{.45}{\includegraphics{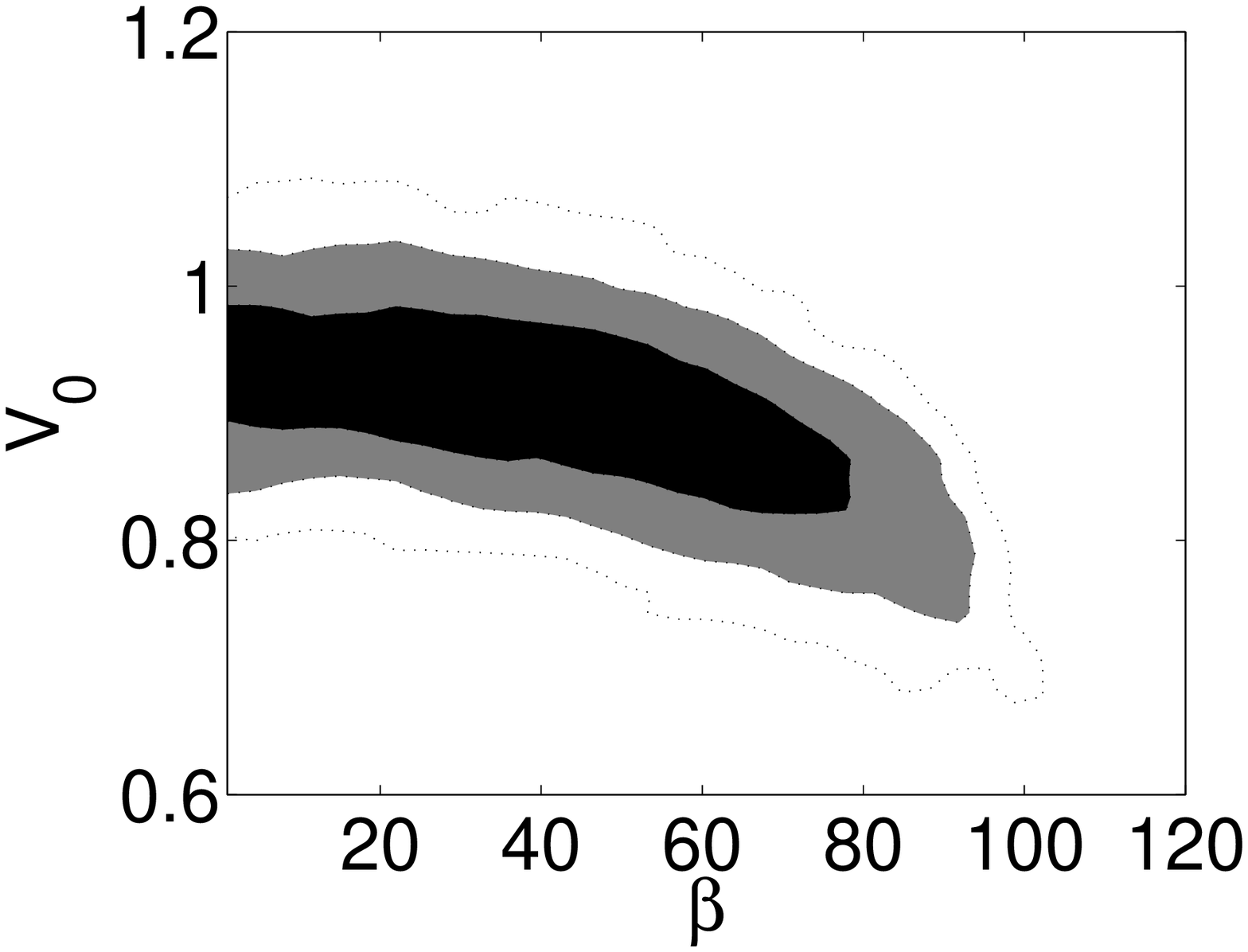}}
}
\centerline{
\scalebox{.45}{\includegraphics{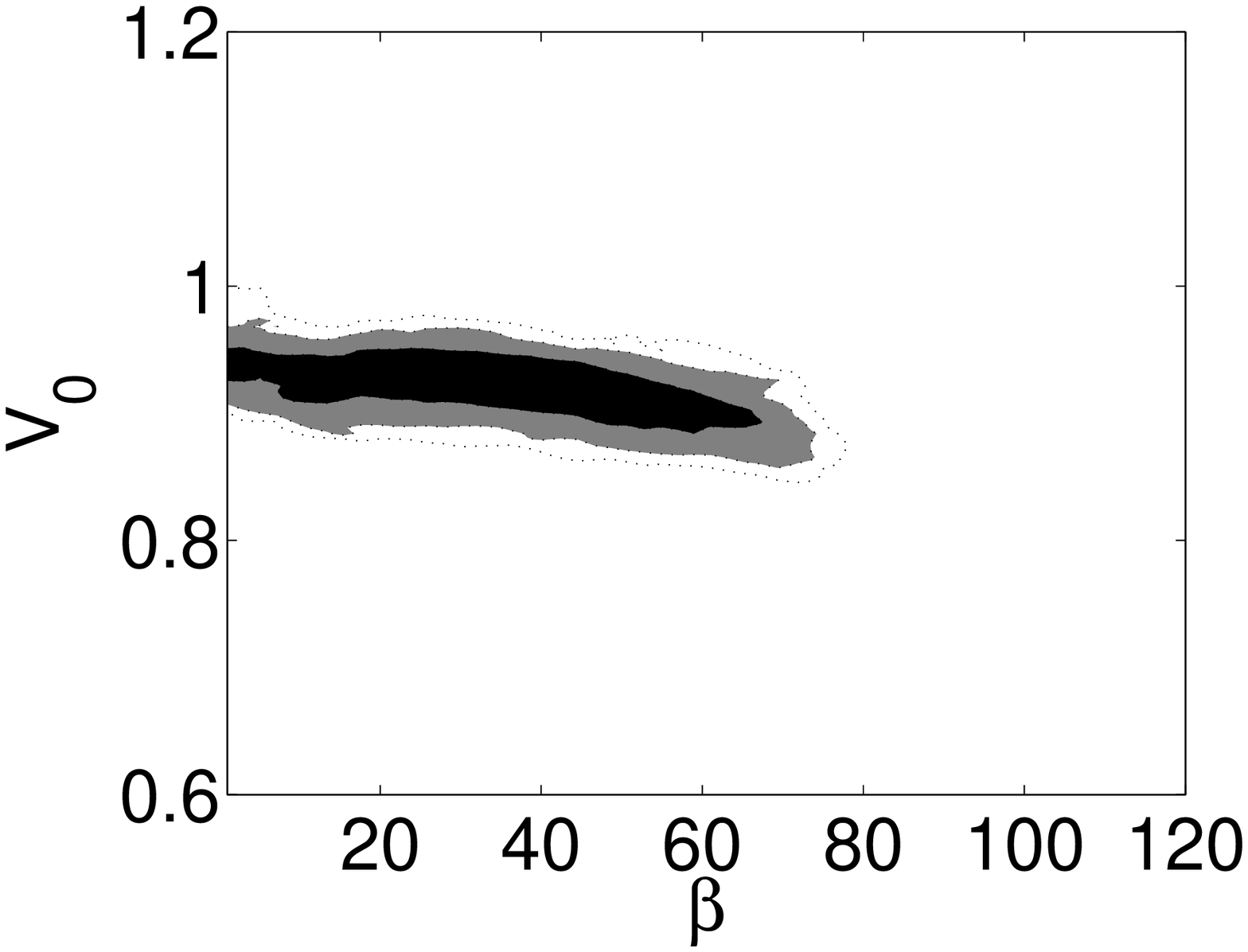}}
}
\caption{Likelihood contours for the AS model in $\beta$ and $V_0$ for
  Stage 2 (top) and Stage 3 photometric, optimistic (bottom), for data
  generated from a cosmological constant model.  Though constants in
  $V_0$ improve greatly, the AS model has little observable difference
  from a cosmological constant for values of $\beta$ below $50$.} 
\label{fig:lambdafid1}
\end{figure}

\begin{figure}
\centerline{
\scalebox{.45}{\includegraphics{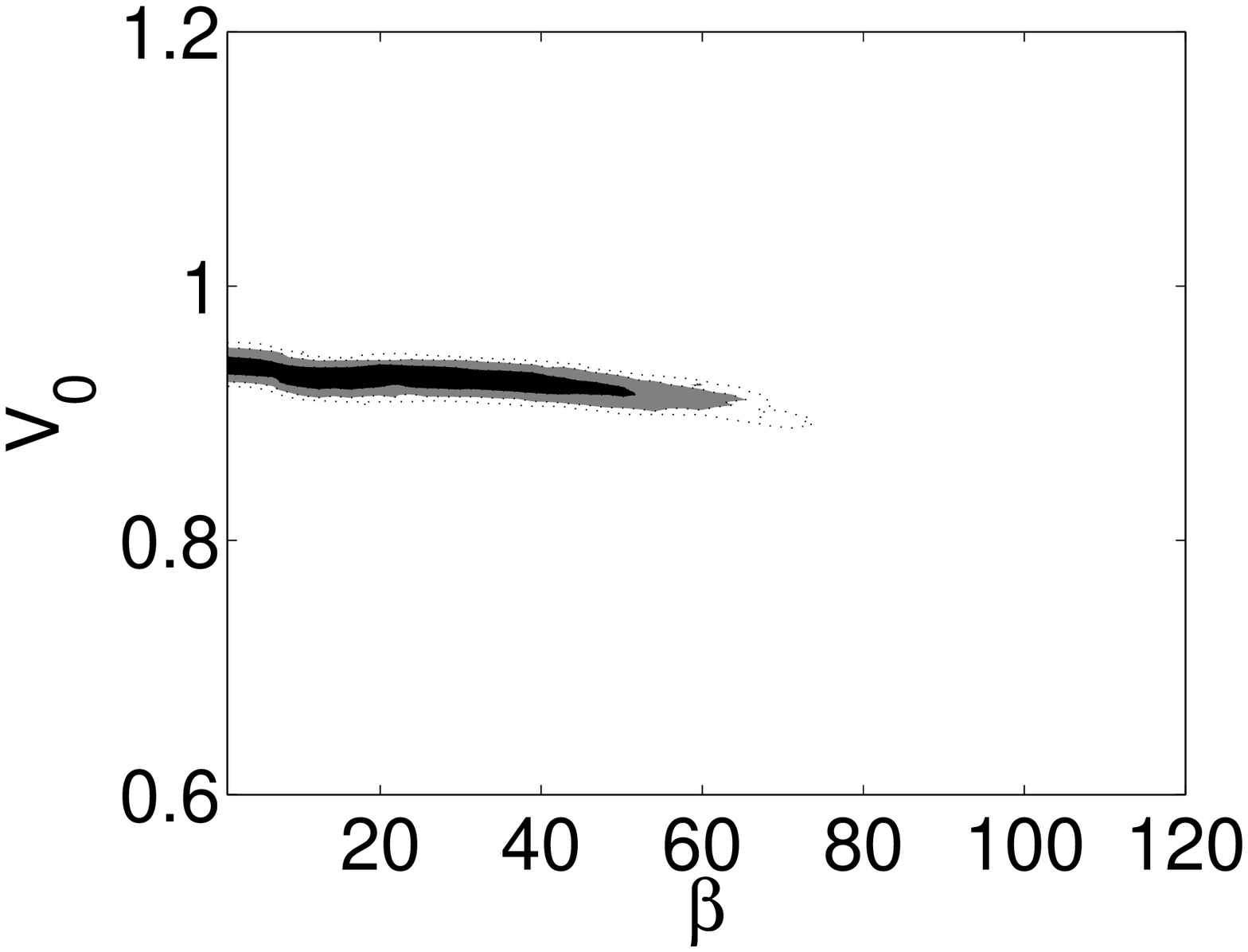}}
}
\centerline{
\scalebox{.45}{\includegraphics{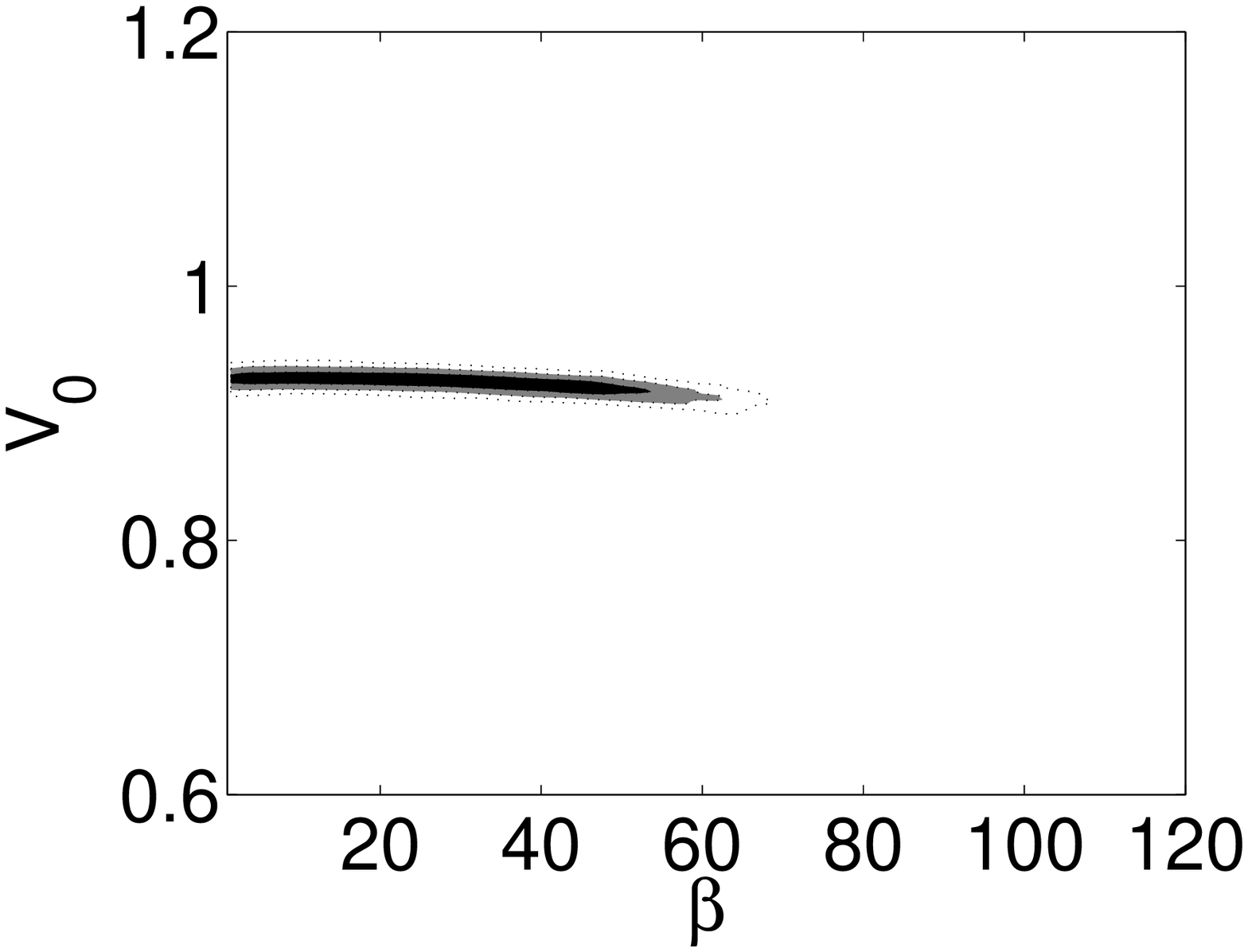}}
}
\caption{Likelihood contours for the AS model in $\beta$ and $V_0$ for
  Stage 4 space (top), and Stage 4 ground LST (bottom), both optimistic
  cases, for data generated from a cosmological constant model.
  Though constants in $V_0$ improve greatly, the AS model has little
  observable difference from a cosmological constant for values of
  $\beta$ below around $50$. The different regions represent $68.27$\%,
  $95.44$\% and $99.73$\% confidence regions.  This convention is used
  for all contour plots in this paper.} 
\label{fig:lambdafid2}
\end{figure}

Having chosen to explore the scalar field parameter space in terms of
the parameters 
$\beta$, $V_0$, and the initial value of the
scalar field (with the other parameters constrained), we then performed an MCMC analysis on simulated data sets generated
around a cosmology with a cosmological constant. The technical details
of this work are presented in the appendix of
\cite{AbrahamseMock}. These data sets are 
consistent with those used by the DETF, using simulated Stage 2, Stage
3, and Stage 4 supernova, weak lensing, baryon oscillation, and CMB
observations.  We did not use the cluster data because of the
difficulty of adapting the DETF construction to a quintessence cosmology,  
nor have we included possible improvements from cross-correlations
\cite{Schneider:2006br,Zhan:2006gi}.
In
the DETF language Stage 2 encompasses projects that are fully funded
and underway, Stage 3 models medium term, medium cost proposals, and
Stage 4 are the larger projects, such as a large ground-based
survey or a new space telescope. 

Though there are three parameters controlling the state of the scalar
field, the initial value of the scalar field is generally washed out
by the tracking behavior, and the variance in height of the minimum is
mostly a reflection of the uncertainty in the Hubble parameter.  This
leaves us with one important parameter, $\beta$, that mostly controls
the unique properties of the AS model.  We include a $V_0$ axis in
the contour plots for convenience. One should take care, though, to
remember that $V_0$ is not a parameter that can be said to have a significant effect on the equation of
state of the dark energy.

Figure \ref{fig:lambdafid1} shows the likelihood contours for DETF Stage 2
and Stage 3 (optimistic) data. From the point of view of these data
sets the behavior of the AS model can be thought of as trending toward
a cosmological constant as $\beta$ goes toward zero.
While these plots show that the constraints on $V_0$, which
corresponds to the dark energy density now, improve greatly with stage
number, the critical consideration here is $\beta$.  The constraints
on $\beta$ do not appear to improve in as dramatic a manner. 
This is because there is a substantial  range of $\beta$ where the
these data sets cannot distinguish the AS model from a cosmological constant.
As seen in Fig \ref{fig:wofacomb}, at $\beta=50$ already one must look
back to $a=.2$ ($z=4$) to see a large deviation from $w=-1$.  To some
extent this feature will affect any attempt to evaluate the AS model
using data from a universe with a cosmological constant.  
In that low $\beta$ region, the small oscillations of the field
persist at late times, and the Stage 4 data sets may be strong enough
to pick up on some effects of these oscillations. Figure
\ref{fig:lambdafid2} shows likelihood contours for DETF Stage 4 space
and ground (optimistic) data. There is a kink in the Stage 4 space
contour plot which corresponds to a range of $\beta$ where the
residual oscillations at late times peak in amplitude.  As the
oscillations represent an energy density above the amount
corresponding to the minimum of the potential, $V_0$ must be lower to
accommodate regions with larger late time amplitudes.

\begin{figure}
\centerline{
\scalebox{.45}{\includegraphics{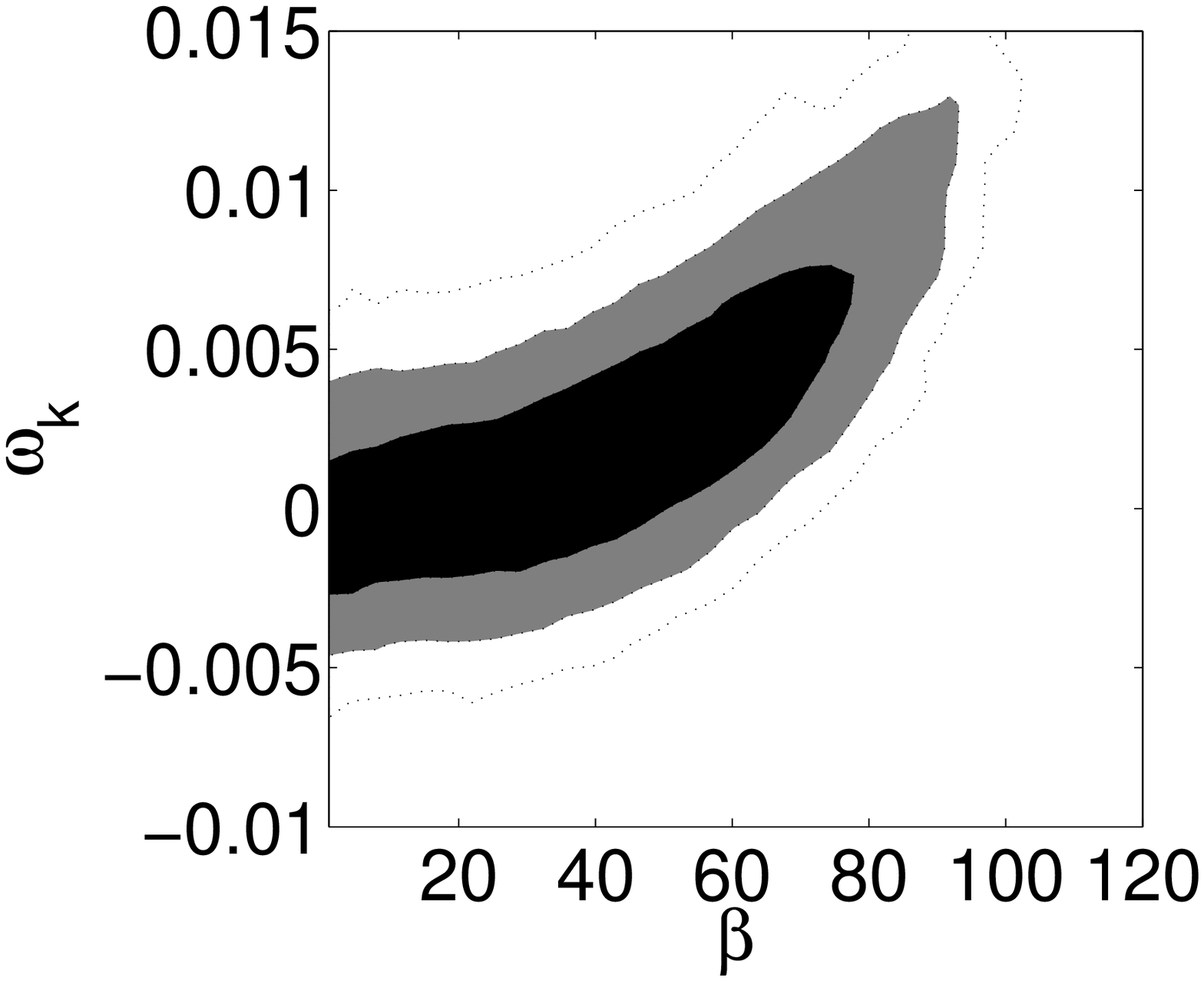}}
}
\centerline{
\scalebox{.45}{\includegraphics{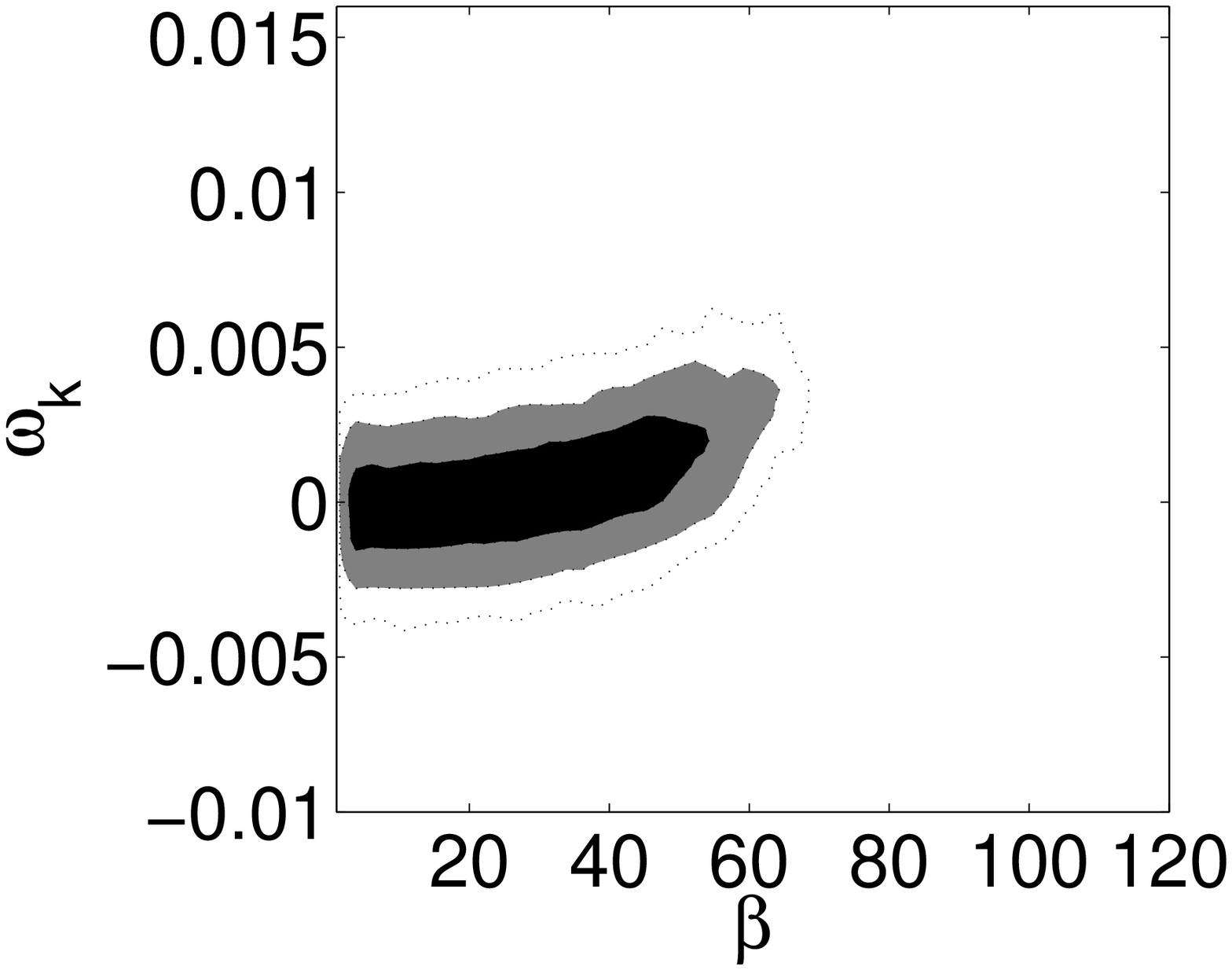}}
}
\caption{This figure illustrates the confusion of the AS model with
  curvature.  The Stage 2 set (top) allow significant departures from
  flatness, even when using data generated by a cosmological constant
  model with no curvature. The Stage 4 (ground, optimistic) data set
  (bottom) are effective at constraining this degeneracy.}  
\label{fig:kcomp}
\end{figure}

Figure \ref{fig:kcomp} shows contours in $\omega_k-\beta$ space
($\omega_k$, the ``curvature density'' is a measure of the curvature of
the universe). There are well known degeneracies associated  with
determining curvature and the dark energy equation of state
simultaneously \cite{knox:023503}.  Here these degeneracies show up by
allowing ``best fits'' to data from a flat universe using non-zero
curvature for certain values of $\beta$, as can be seen in contours
for the Stage 2 data.  The Stage 4 data set shown in the lower panel clamps down on this
behavior considerably.

\section{MCMC of the AS model around AS model-based data sets}

\begin{figure}[ht]
\centerline{
\scalebox{.45}{\includegraphics{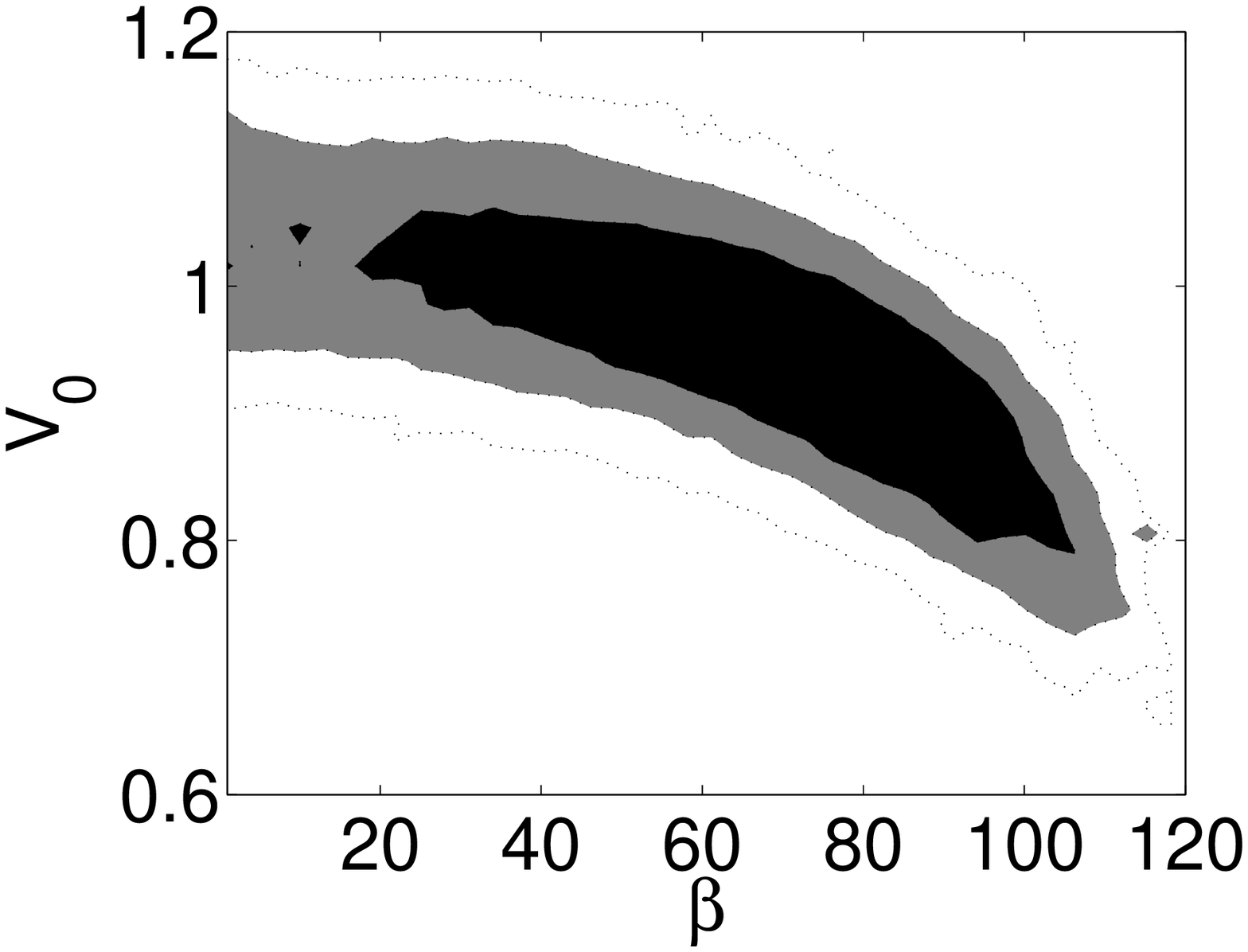}}
}
\centerline{
\scalebox{.45}{\includegraphics{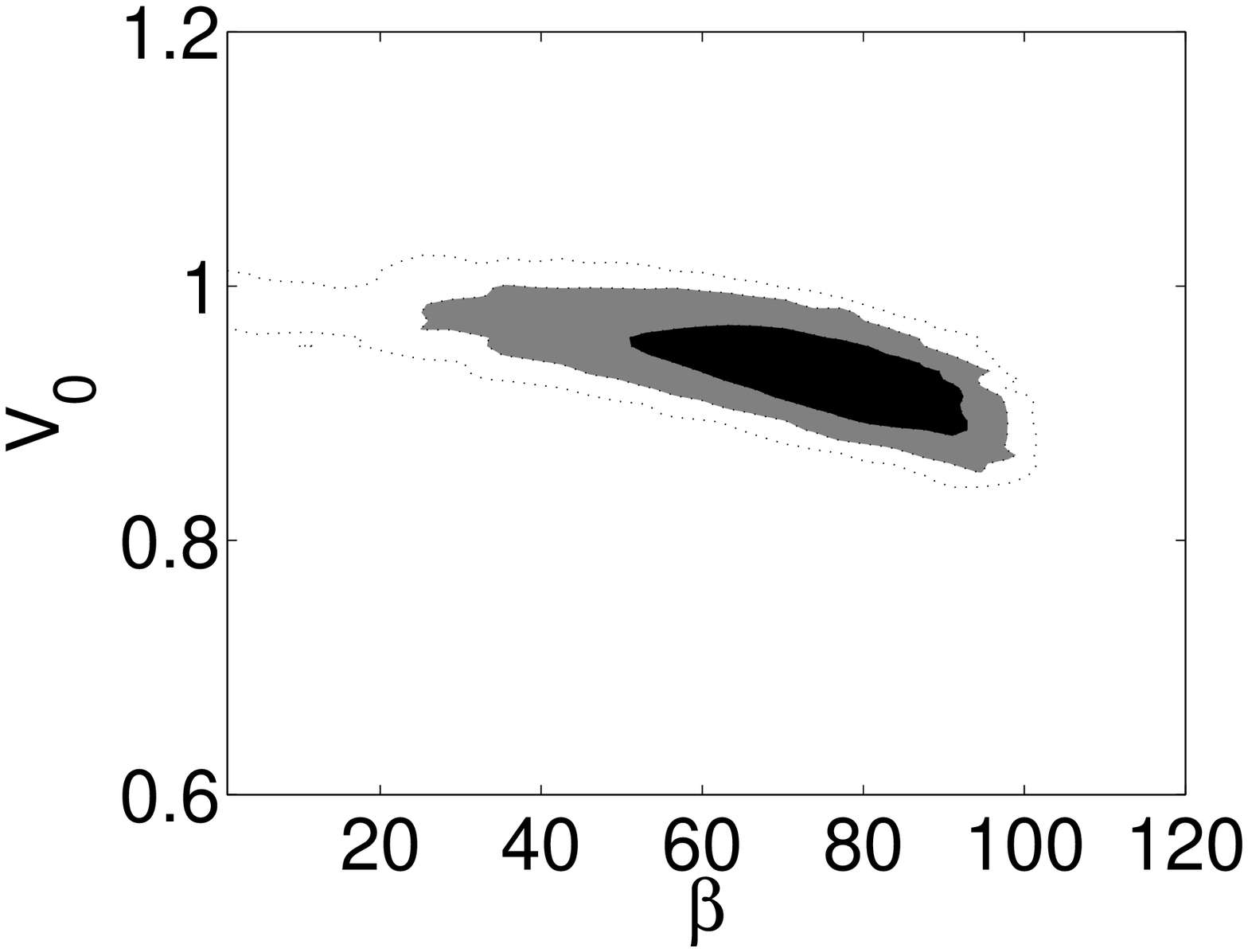}}
}
\caption{Likelihood contours for the AS model in $\beta$ and $V_0$ for Stage 2 (top) and Stage 3 photometric, optimistic (bottom), for data generated from the $\beta=80$ AS model.}
\label{fig:ASefid1}
\end{figure}

\begin{figure}[ht]
\centerline{
\scalebox{.45}{\includegraphics{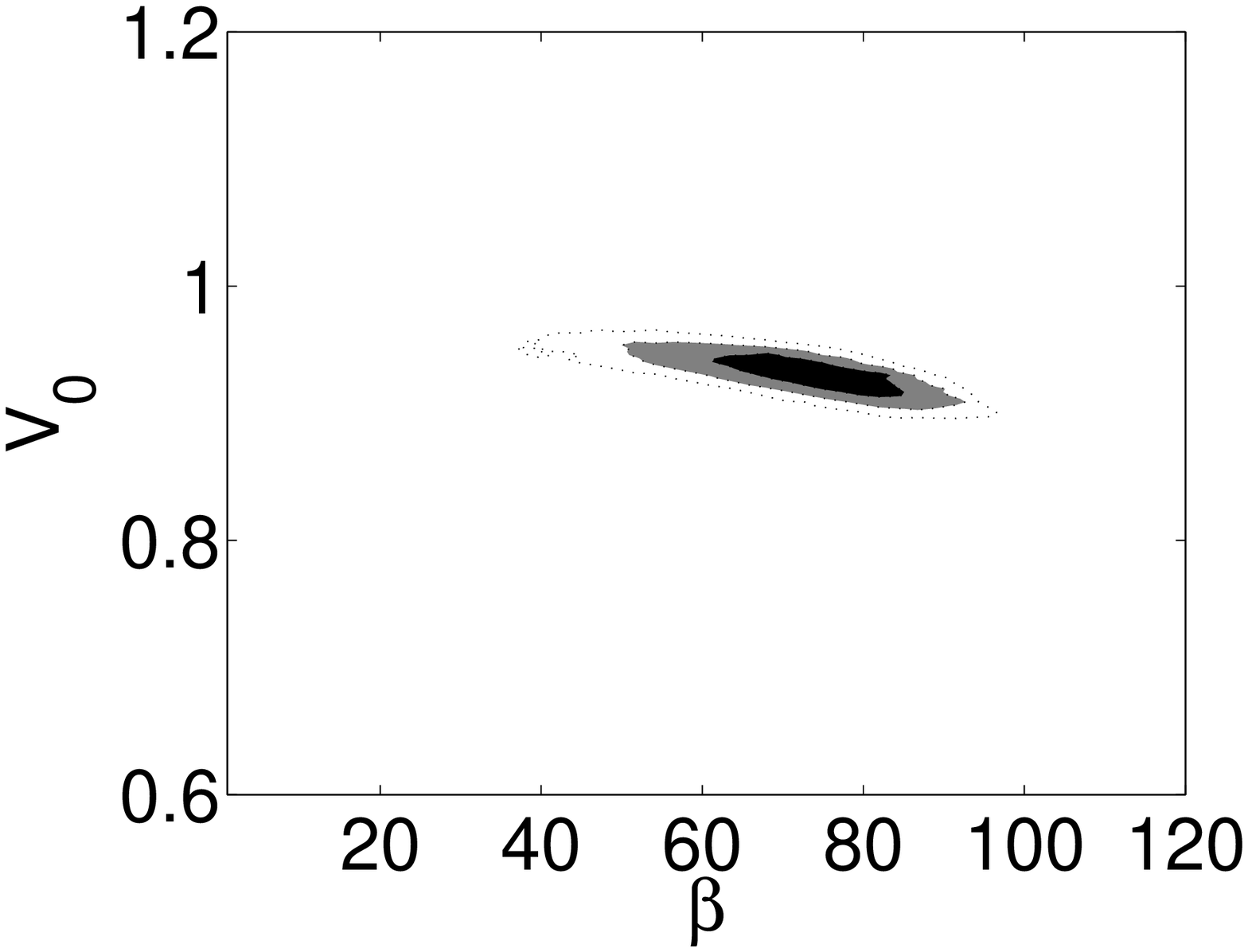}}
}
\centerline{
\scalebox{.45}{\includegraphics{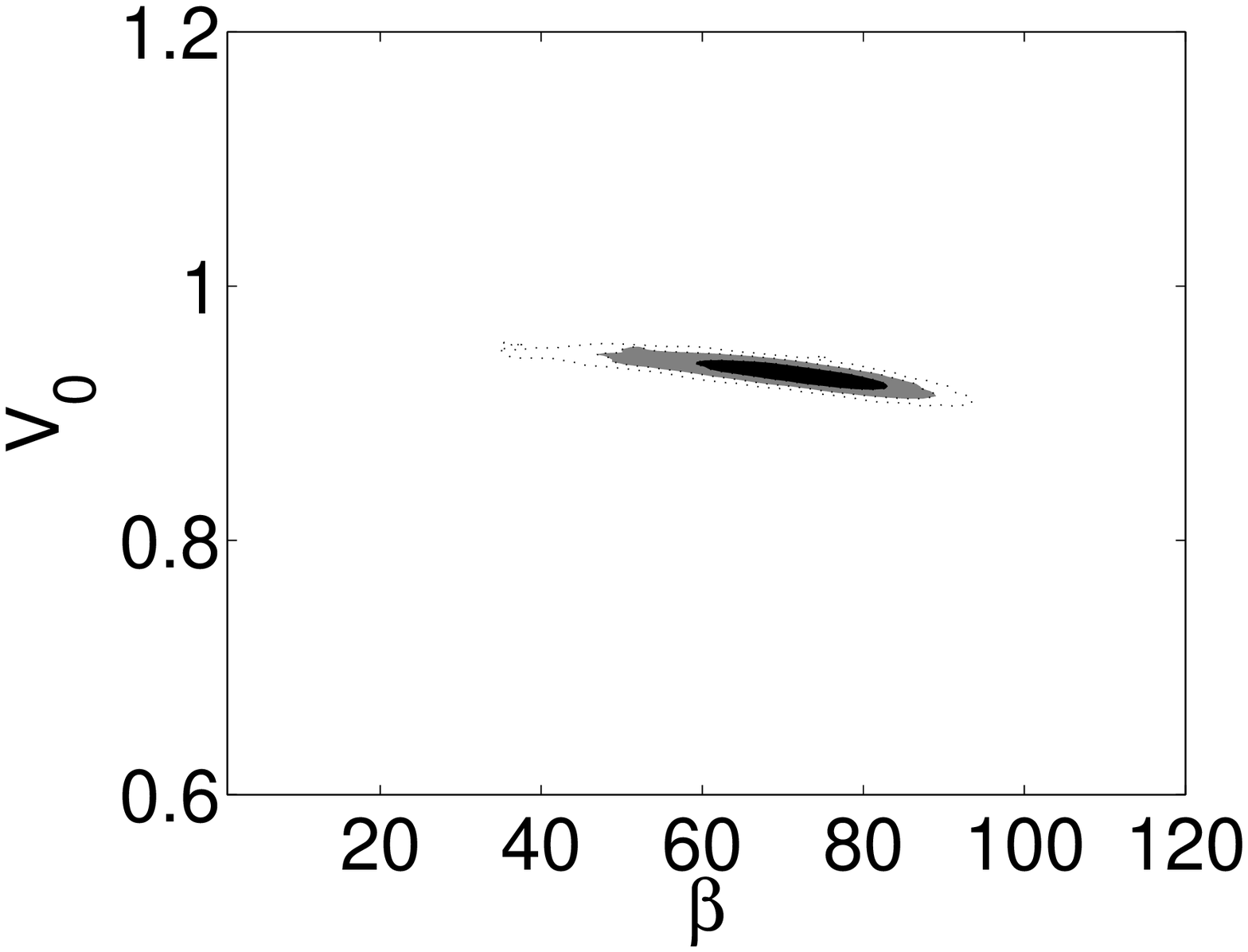}}
}
\caption{Likelihood contours for the AS model in $\beta$ and $V_0$ for
  Stage 4 space (top), and Stage 4 ground LST, both optimistic (bottom)
  cases, for data generated from the selected AS model.} 
\label{fig:ASefid2}
\end{figure}

As a way of further exploring the power of the advanced data sets, a
sample AS model was chosen as a new fiducial model around which the
data sets were generated.  This model was chosen to be marginally
consistent with the cosmological constant-based Stage 2 data, but different enough from a
cosmological constant that a cosmological constant would be strongly
ruled out with Stage 4 data (the point being to illustrate the power
of Stage 4). The fiducial model chosen here has a $\beta=80$, as 
depicted in Fig. \ref{fig:wofacomb}.
Because the Stage 2 data allowed larger values of $\beta$ when
curvature was present, the sample AS model was
chosen with non-zero spatial curvature to make it more consistent with
a flat cosmological constant model for this value of $\beta$.
The results of the calculation, as seen in Fig. \ref{fig:ASefid1} and \ref{fig:ASefid2},
show that, for a universe described by this specific model, the Stage
4 experiments will rule out a cosmological constant by several sigma.

\section{Discussion and Conclusions}
We have analyzed the impact of the DETF simulated data sets in the
context of the Albrecht-Skordis scalar field model of dark energy. We
find that the effect of the DETF data sets on the parameter space 
of the AS model is very similar to the DETF findings.  There is
substantial improvement in the constraining power of each successive
stage of experiments.  We presented likelihood contour plots in AS
parameter space for a few key combinations of DETF simulated data. We
used these plots to demonstrate the broad agreement with the DETF
results and also to point out more subtle effects such as
degeneracies between the dark energy equation of state 
and curvature and some peculiarities of the way the AS model can
approximate a cosmological constant. In the course of this work we
have produced similar contour plots for many more instances of the
DETF simulated data (taking individual techniques separately for
example, and using both the DETF optimistic and pessimistic data
models).  We find our overall conclusions and detailed points apply
quite generally across the full range of DETF simulated data. 
We also find that as one moves to better experiments, the improved
discriminating power will definitely rule out interesting portions of
the AS model space and could completely rule out the cosmological
constant if the AS model is correct.  

The DETF reported a figure of merit in terms of the inverse of the
area inside of likelihood contours in their model space, $w_0$ and
$w_a$.  This figure of merit showed an improvement of a factor of
three going from Stage 2 to Stage 3, and a factor of ten going from
Stage 2 to Stage 4, assuming good Stage 3 and Stage 4 projects. In the
parameter space we used for the AS model we have only one parameter
that strongly affects the equation of state.  The constraints on this parameter, $\beta$,
show improvement by roughly the square roots of these factors,
appropriate for the reduced dimensionality in the AS parameter
space. We also see some subtle differences between the
constraining power of ground and space Stage 4 data models 
in that the other variable plotted, $V_0$, is somewhat more
strongly constrained by the Stage 4 ground data set than for the Stage
4 space. Other scalar field models \cite{AbrahamseMock,BozekMock} display
the opposite behavior.  We are currently investigating this effect
further, which may lead to interesting insights into the
complementarity of ground and space-based Stage 4 experiments. 

It is interesting to consider the relationship between this work and
recent work by one of us and G. Bernstein \cite{Albrecht:103003}.
There the same DETF data sets were studied in the context of an
abstract dark energy model where the equation of state was modeled by
many more than the two parameters used by the DETF.  One conclusion of
\cite{Albrecht:103003} was that high quality data can make good
measurements of significantly more than two equation of state
parameters.  We have seen in this paper that the main equation of state
parameter $\beta$ of the AS model is constrained by DETF data sets to a
similar degree as the DETF $w_o-w_a$ parameters, even though they
describe very different functions $w(a)$. 
As discussed in \cite{Albrecht:2007xq},
we believe that this ability
to constrain a wide variety of functions $w(a)$ is another
manifestation of the rich constraining power demonstrated in
\cite{Albrecht:103003}.

A good way to
think about this may be to consider the expansion of the AS family of
$w(a)$ function as well as the $w_0-w_a$ family of functions in terms
of the independently measured orthogonal functions 
$w_i(a)$ from \cite{Albrecht:103003}. The fact that we have seen here
(and also in \cite{AbrahamseMock,BozekMock}) that a variety of
different $w(a)$ functions can be constrained as well as the DETF $w_0-w_a$
functions appears to reflect the fact that fundamentally many more
functions $w(a)$ are measured than are contained in any one of these
families alone. 

An upshot of this and the companion work in
\cite{AbrahamseMock,BozekMock}) is that modeling the impact of future
experiments using the DETF parameters appears to give a good indication of the
impact on quintessence dark energy models with a similar number of
parameters.  An advantage of the methods used here is that one can see
explicitly how future data can constrain real dark energy models in a
significant way, and can even eliminate some models entirely. 

\begin{acknowledgments}
We wish to acknowledge
Lloyd Knox,
Jason Dick,
and
Michael Schneider
for conversations and consultation that contributed to this paper.
We would also like to thank
David Ring for finding an error in our code,
and Tony Tyson and his group (especially Perry Gee and Hu Zhan) for use of their
computational resources. We thank Gary Bernstein for providing us with
Fischer matrices suitable for adapting the DETF weak lensing data
models to our methods. This work was supported in part by DOE grant
DE-FG03-91ER40674 and NSF grant AST-0632901. 
\end{acknowledgments}
\bibliography{bib01}
\end{document}